\documentclass[
    ,final            % use final for the camera ready runs
  ]
  {aipproc}

\layoutstyle{6x9}

\def\lya{\ifmmode {\rm Ly}\alpha~ \else Ly$\alpha$~\fi}
\def\lyb{\ifmmode {\rm Ly}\beta~ \else Ly$\beta$~\fi}
\def\lyg{\ifmmode {\rm Ly}\gamma~ \else Ly$\gamma$~\fi}
\def\civ{\ifmmode {\rm C}\,{\sc iv}~ \else C\,{\sc iv}~\fi}
\def\civn{\ifmmode {\rm C}\,{\sc iv} \else C\,{\sc iv}\fi}
\def\cvn{\ifmmode {\rm C}\,{\sc v}~ \else C\,{\sc v}~\fi}
\def\cvin{\ifmmode {\rm C}\,{\sc vi}~ \else C\,{\sc vi}\fi}
\def\nvn{\ifmmode {\rm N}\,{\sc v} \else N\,{\sc v}\fi}
\def\nv{\ifmmode {\rm N}\,{\sc v}~ \else N\,{\sc v}~\fi}
\def\nvin{\ifmmode {\rm N}\,{\sc vi}~ \else N\,{\sc vi}~\fi}
\def\nviin{\ifmmode {\rm N}\,{\sc vii} \else N\,{\sc vii}\fi}

\def\neix{\ifmmode {\rm Ne}\,{\sc ix}~ \else Ne\,{\sc ix}~\fi}
\def\nex{\ifmmode {\rm Ne}\,{\sc x}~ \else Ne\,{\sc x}~\fi}
\def\hi{\ifmmode {\rm H}\,{\sc i}~ \else H\,{\sc i}~\fi}

\def\chandra {{\it Chandra}~}

\def\xmmn {{\it XMM-Newton}}

\begin{document}

\title{AGN Feedback: Does it work?}


\author{Smita Mathur}{
  address={The Ohio State University}
}

\author{Rebecca Stoll}{
  address={The Ohio State University}
}

\author{Yair Krongold}{
  address={Universidad Nacional Autonoma de Mexico}
  %,altaddress={<author1 address>} % additional visiting address
}

\author{Fabrizio Nicastro}{
  address={Harvard-Smithsonian Center for Astrophysics}
}
\author{Nancy Brickhouse}{
  address={Harvard-Smithsonian Center for Astrophysics}
}
\author{Martin Elvis}{
  address={Harvard-Smithsonian Center for Astrophysics}
}

\begin{abstract}
While feedback is important in theoretical models, we do not really know
if it works in reality. Feedback from jets appears to be sufficient to
keep the cooling flows in clusters from cooling too much and it may be
sufficient to regulate black hole growth in dominant cluster
galaxies. Only about 10\% of all quasars, however, have powerful radio
jets, so jet-related feedback cannot be generic.  The outflows could
potentially be a more common form of AGN feedback, but measuring mass
and energy outflow rates is a challenging task, the main unknown being
the location and geometry of the absorbing medium. Using a novel
technique, we made first such measurement in NGC 4051 using XMM data and
found the mass and energy outflow rates to be 4 to 5 orders of magnitude
{\it below} those required for efficient feedback. To test whether the
outflow velocity in NGC 4051 is unusually low, we compared the ratio of
outflow velocity to escape velocity in a sample of AGNs and found it to
be generally less than one. It is thus possible that in most Seyferts the 
feedback is not sufficient and may not be necessary. 
\end{abstract}

\maketitle

%%%%%%%%%%%%%%%%%%%%%%%%%%%%%%%%%%%%%%%%%%%%
%% MAINMATTER
%%%%%%%%%%%%%%%%%%%%%%%%%%%%%%%%%%%%%%%%%%%%
%\section{}
\subsection{AGN outflows}

Feedback from AGNs has been invoked to solve a number of astrophysical
problems, ranging from cluster cooling flows to the structure of
galaxies.  The outflows are ubiquitous in AGNs and could potentially
provide a form of AGN feedback more common than jets; they may also be
responsible for enriching the intergalactic medium with
metals. Understanding the physical conditions in the absorbing outflows,
particularly measuring their mass and energy outflow rates, thus becomes
very important.

 With the gratings available on
\chandra and \xmmn, the past 10 years have seen a great progress in
understanding the physical properties and kinematics of warm-absorber
outflows\footnote{There have been several claims of relativistic
outflows in the literature; we do not discuss them here. Vaughan \&
Uttley (2008) have argued that these are consistent with statistical
fluctuations, given publication bias.}. A consistent picture has emerged
from recent observations of several AGNs. The absorbers have at least
two components: with a low ionization parameter (LIP) and a high
ionization parameter (HIP). The LIP and HIP phases appear to be in
pressure equilibrium, and so likely emerge from the common wind
(e.g. Netzer et al. 2003). The LIP component is also responsible for the
UV absorption lines (often showing multiple components unresolved in
X-ray), but the HIP is not. Several AGNs, especially those observed with
the high energy transmission grating (HETG) on \chandra, show signs of
an additional super-high ionization component (SHIP), also in pressure
equilibrium with other components. In spite of these strides, there are
still important unanswered questions: where do the outflows originate
from; an accretion disk (e.g. as suggest by Proga, 2007) or the obscuring
torus as suggested by Krolik \& Kriss (1995)? This
uncertainty of six orders of magnitude in radial distance is
embarrassing. As noted above, an important question, particularly for
the feedback process, is how much mass, energy and momentum do they
carry.

\subsection{Feedback from outflows: a challenging measurement}

\begin{figure}[t]
  \includegraphics[height=.34\textheight]{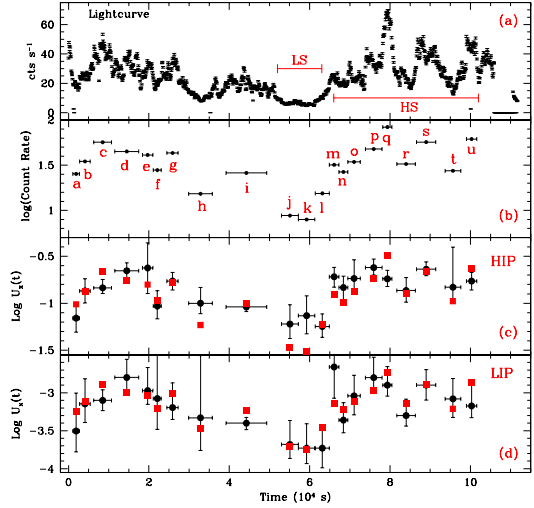}
  \caption{The XMM light-curves of NGC\,4051. The continuum and the
 binned continuum are shown in the top two panels. The bottom two panels
 show variations in the ionization parameter of HIP and LIP. For
 comparison with the continuum level, the red squares represent the
 count rate with a constant offset. Note that the HIP is is
 out-of-equilibrium at times h, j, k, q.}
\end{figure}

  In theory, energy injection efficiencies range from order of unity
($L_{outflow}/L_{bolometric}\approx 1$) to a minimum of 5\% (e.g. Silk,
2005 and Scannapieco \& Oh, 2004).  However, do we know whether actual
quasar outflows can indeed carry such energy?  This is a challenging
measurement to make, the main unknown being the location and geometry of
the absorbing medium. If the absorber geometry is, for example, a thin
shell of gas located far from the nuclear black hole, then for a given
column density the implied mass could be quite large compared to an
absorber located closer in. The mass together with the outflow velocity
(measured from the blueshifts of absorption lines) then allows us to
calculate the energy outflow rate. Thus it is vital to know the location
of the absorber.

 However, in the equation for the photoionization parameter
($U\propto L/ n_e R^2$), the radius of the absorbing region ($R$) is
degenerate with the density ($n_e$). This degeneracy can be broken if we
can determine the density independently. Since the recombination times
are inversely proportional to density, the response of absorption lines
to continuum variations during the ionizing phase provides a robust
density diagnostic.  Thus we must probe the appropriate time domain for
an ionizing/recombining (rather than photoionization-equilibrium) plasma. 

%Why is it so difficult to measure densities in absorbing gas using
%variability analysis? In X-rays, the integration times have been too
%long to make meaningful time series analysis.  In the UV, there have
%been several monitoring campaigns, but with their sparse time sampling,
%they could only put lower limits on density and thus upper limits on the
%radius of the absorber and often these limits were uninteresting in
%localizing the outflow radius. The breakthrough finally came with the
%XMM observation of NGC\,4051.

\subsection{The outflow rate in NGC\,4051}

\begin{figure}[t]
  \includegraphics[height=.385\textheight]{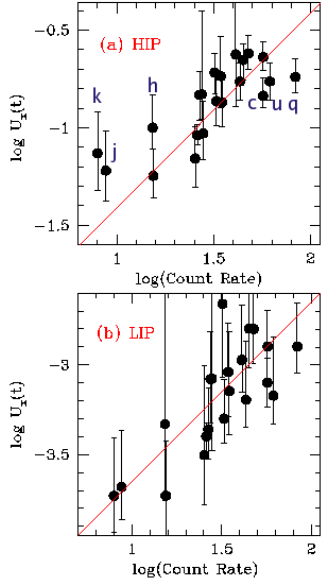}
  \caption{The ionization parameters of HIP (top) and LIP (bottom) are
 plotted as a function of the count rate. Both components follow the
 photoionization equilibrium, except for the points h, j, k, q of the HIP.}
\end{figure}

In an XMM observation, a high resolution grating spectrum could be
observed simultaneously with the low resolution CCD observation. For
NGC\,4051, the grating spectrum provided the accurate baseline model of
the absorber while the time resolved spectroscopy could be performed
with the CCD data. Because of the broad features in the spectrum, such
as the Fe UTAs (unresolved transition arrays), the changes in the
ionization state of the absorbers could be tracked even with the CCD
data, knowing the base-line model (see figures 1, 2).

Using this novel technique, Krongold et al. (2007) managed
to determine the absorber density in this system, so the distance.  They
determined the distance of the HIP from the nucleus to be 1\,light day,
and that of the LIP to be $\le 3.5$\,light days. These results strongly
argue in favor of an accretion disk origin for the winds.  For the
bi-conical outflow geometry expected from disks, the resulting mass
outflow is $\sim10^{-4}$\,M$_{\odot}$\,yr$^{-1}$ with an energy outflow
rate of $\sim10^{38}$\,erg s$^{-1}$, 4 to 5 orders of magnitude {\it
below} those required for efficient feedback.

\subsection{Scaling the outflows in other AGNs}

How generic is the NGC\,4051 result? NGC\,4051 may be an unusual object,
given that it is a narrow-line Seyfert 1, has low luminosity, and a low
mass black hole. In NGC\,4051, the outflow velocity is a small fraction
of the escape velocity at the wind launching radius. Perhaps we are
observing the wind before it is fully accelerated in this AGN because of
the presence of a transverse flow in our line of sight. In other AGNs,
perhaps, v$_{wind}$ is equal to v$_{escape}$. To test whether this is
indeed the case, we compared v$_{wind}$ to v$_{escape}$ for a sample of
AGNs. Here v$_{wind}$ is the outflow velocity as observed in UV
absorption lines and v$_{escape}$ was calculated scaling the wind radius
from the NGC\,4051 value (see Stoll et al. 2009 for details). 

As shown in figure 3, for most AGNs v$_{wind}/$v$_{escape}\ll 1$. The
result remains the same if we scale the wind radius with L$^{0.5}$ or
with black hole mass. Thus NGC\,4051 does not appear to be unusual in
its outflow velocity. It is thus highly likely that mass and energy
outflow rate in most Seyfert galaxies is much too small to account for
the kind of feedback required by theoretical models (see Stoll et
al. 2009 for possible caveats).

\begin{figure}[t]
  \includegraphics[height=.3\textheight]{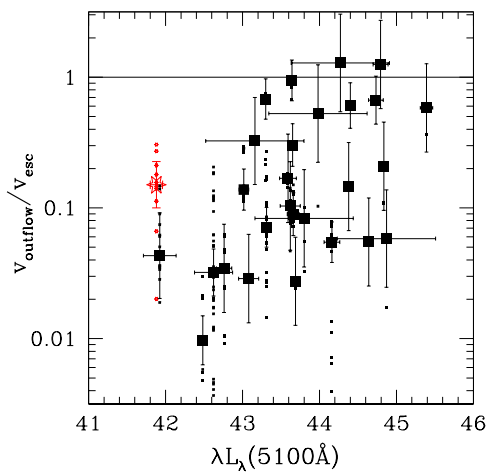} 
   \caption{The ratio of outflow velocity to the escape velocity appears
  to be less than one in most AGNs. The red diamond point is for NGC\,4051.}
\end{figure}




%\bibliography{sample}


%%%%%%%%%%%%%%%%%%%%%%%%%%%%%%%%%%%%%%%%%%%
%% The following lines show an example how to produce a bibliography
%% without the help of the BibTeX program. This could be used instead
%% of the above.
%%%%%%%%%%%%%%%%%%%%%%%%%%%%%%%%%%%%%%%%%%%

\end{document}